\documentclass[11pt,reqno]{amsart}

\newtheorem{thm}{THEOREM}

\newtheorem{lem}{LEMMA}

\theoremstyle{definition}
\newtheorem{rem}{REMARK}

\newcommand{\infspec}{{\rm inf\ spec\ }}
\newcommand{\R}{{\mathbb R}}

\renewcommand{\d}{\cdot}

\newcommand{\C}{{\mathbb C}}
\newcommand{\ap}{ \alpha\pi^{-1}}

\newcommand{\Da}{{D}^\ast}

\newcommand{\F}{{\mathcal F}}
\newcommand{\Em}{{\mathcal E}}
\newcommand{\Q}{{\mathcal Q}}
\newcommand{\Ll}{{\mathcal L}}

\newcommand{\Hh}{{\mathcal H}}

\newcommand{\aan}{a_{\lambda}}
\newcommand{\ac}{a^{\ast}_{\lambda}}

\newcommand{\ean}{\varepsilon_{\lambda}}
\newcommand{\half}{\mbox{$\frac{1}{2}$}}

\newcommand{\al}{{\alpha}}
\newcommand{\pa}{{\parallel}}

\newcommand{\so}{{\Sigma_\al}}

\newcommand{\po}{{\Psi_0}}

\newcommand{\T}{T_0}

\newcommand{\as}{\sqrt{\alpha}}

\newcommand{\Ow}{{\mathcal O}}
\newcommand{\ora}{{|0\rangle}}

\newcommand{\la}{\Lambda}

\newcommand{\const}{{\rm const.}}
\newcommand{\mA}{{\mathcal A}}

\newcommand{\pe}{\psi_{n+1}}

\newcommand{\be}{\beta}

\begin{document}

\title[Enhanced Binding in non-relativistic QED]{Enhanced binding in non-relativistic QED}
\author[Hainzl]{Christian Hainzl$^1$}
\address{Mathematisches Institut, LMU M\"unchen,
Theresienstrasse 39, 80333 Munich, Germany}
\email{hainzl@mathematik.uni-muenchen.de}

\author[Vougalter]{Vitali Vougalter}
\address{ Department of Mathematics, University of British Columbia,
Vancouver, B.C. V6T 1Z2, Canada}
\email{vitali@math.ubc.ca}

\author[Vugalter]{Semjon A. Vugalter}
\address{Mathematisches Institut, LMU M\"unchen,
Theresienstrasse 39, 80333 Munich, Germany}
\email{wugalter@mathematik.uni-muenchen.de}

\date{\today}

\begin{abstract}
We consider a spinless particle coupled to a photon field
and prove that even if the Schr\"odinger operator $p^2 + V$ does not
have eigenvalues the system can have a ground state. We describe the coupling by means of the
Pauli-Fierz Hamiltonian and our
result holds in the case where the coupling constant $\al$ is small.
\end{abstract}

\maketitle

\footnotetext[1]{Marie Curie Fellow}

\section{INTRODUCTION}

In the picture of Quantum electrodynamics (QED) atoms consist of
char\-ged particles, which are necessarily coupled to a photon
field. If one neglects the radiation effects one obtains the standard
Schr\"odinger operator. Although the fundamental properties of the
one-particle and multi-particle Schr\"odinger operators have been
successfully studied since the middle of the last century, the
systematic mathematical study of the non-relativistic QED model
was initiated by Bach, Fr\"ohlich, and Sigal  in
\cite{BFS1,BFS2,BFS3} only a couple of years ago (a comprehensive
review of results in non-relativistic QED can be found in
\cite{GLL}) and some very fundamental problems remain still
open. One of these problems is the question of enhanced binding
via interaction with a quantized radiation field.

Consider a particle in a potential well $\be
V(x)$ with $V(x) \leq 0$. If the potential well is not deep
enough, i.e. $\be$ is small, the corresponding Schr\"odinger
operator does not have a discrete spectrum and binding does not
occur. There exists a critical value
$\be_0$ such that for $\be > \be_0$ there is at least one bound
state whereas for $\be \leq \be_0$ no particle can be bound.

In a recent paper Griesemer, Lieb, and Loss (\cite{GLL}) proved
that a photon field cannot decrease the binding energy. If the
Schr\"odinger operator with potential $\be V$ has an eigenvalue,
the corresponding energy operator in non-relativistic QED
(Pauli-Fierz Hamiltonian) has a ground state.

However, the physical intuition tells us that interaction with a
photon field must increase binding. According to the photon cloud
surrounding the particle, the effective mass of the electron 
increases and consequently it
needs more energy to leave the potential well.

The goal of this paper is to give a mathematical rigorous proof
this phenomenon. Previously the enhanced binding was studied in
the dipole approximation by Hiroshima and Spohn in \cite{HS}. In
this approximation it is assumed that the magnetic vector
potential does not depend on the coordinates of the particle. They
proved that, if the potential $\be V$ is fixed, for sufficiently
large values of the coupling parameter $\al$ (which is the fine
structure constant, see (\ref{rpf})), binding takes place.

Our approach to this problem is different. On the one hand we
study the Pauli-Fierz operator without additional restrictions on
the magnetic vector potential and on the other hand our results
hold for small values of $\al$ (recall, that the physical value of
$\al$ is about $1/137$). We prove that in case of the Pauli-Fierz
operator (and for small $\al$ enough) binding starts at values of
$\be$ strictly less than $\be_0$.

\section{MAIN RESULTS}

We describe the self-energy of the particle by
\begin{equation}\label{rpf}
T = (p + \sqrt{\al}A(x))^2 +  H_f,
\end{equation}
acting on the Hilbert space 
\begin{equation}
\Hh = \Ll^{2}({\mathbb {R}}^{3})\otimes \F,
\end{equation}
where $\F$ is the symmetric Fock space for the photon field.

We use units such that  $\hbar=c=1$ and the mass $m=\half$. The
electron charge is then given by $e=\as$, with $\alpha\approx
1/137$ the fine structure constant.
In the present paper $\al$
plays the role of a small, dimensionless number. Our results hold
for sufficiently small values of $\al$. 
The electron momentum operator is $p= -i\nabla_x$, while $A$ is
the magnetic vector potential. We fix the Coulomb gauge ${\rm div}
A =0$.

The vector potential is
\begin{equation}
A(x) = \sum_{\lambda = 1,2} \int_{\R^3}
\frac{\chi(|k|)}{2\pi|k|^{1/2}} \ean \big [\aan(k) e^{ikx} +
\ac(k) e^{-ikx}\big] dk,
\end{equation}
where the operators $\aan, \ac$ satisfy the usual commutation
relations
\begin{equation}
[a_\nu (k),\ac(q)] = \delta(k-q)\delta_{\lambda,\nu}, \quad
[\aan(k), a_\nu(q)] = 0. \ \
\end{equation}
The vectors $\ean(k) \in \R^3$ are two orthonormal
polarization vectors perpendicular to $k$.

Obviously,
\begin{equation}
A(x) = D(x) + D^\ast (x),
\end{equation}
where
\begin{equation}
D(x) = \sum_{\lambda = 1,2} \int_{\R^3}
\frac{\chi(|k|)}{2\pi|k|^{1/2}} \ean \aan(k) e^{ikx}  dk
=\sum_{\lambda = 1,2} \int_{\R^3} G^{\lambda}(k)
a_\lambda(k)e^{ikx}  dk,
\end{equation}
and $D^\ast$ is the  operator adjoint to $D$.

The function $\chi(|k|)$ describes the ultraviolet cutoff for the interaction
at large 
wavenumbers $k$. For convenience we choose  $\chi$  to be the
Heaviside function $\Theta(\Lambda -|k|/l_C)$, where $l_C=
\hbar/(mc)$  is the Compton wavelength. In our units $l_C = 2$.
Our  proof would work for any other cut-off.

The photon field energy $H_f$ is given by
\begin{equation}
H_f = \sum_{\lambda= 1,2} \int_{\R^3} |k| \ac (k) \aan (k) dk,
\end{equation}
whereas
\begin{equation}
P_f =\sum_{\lambda=1,2} \int_{\R^3} k a_\lambda^*(k) a_\lambda(k) dk
\end{equation}
denotes the field momentum.

To prove existence of enhanced binding in non-relativistic QED we
would like to compare binding in the presence of the photon field
and without it. To this end let us introduce the  Schr\"odinger
operator
\begin{equation}\label{sch}
h_\be = -\Delta  + \be V(x)
\end{equation}
with external potential $\be V(x) \in C(\R^3)$, which we assume to
be radial $V(x)=V(|x|)$, non-positive, $V(x) \leq 0$, and with
compact support. It is known that there is a critical value of the
parameter $\be_0 > 0$ such that for $\be \leq \be_0$ there is no
ground state and the operator (\ref{sch}) has only an essential
spectrum and at the same time for all $\be > \be_0$ the operator
$h_\be$ has at least one eigenvalue.

The corresponding operator with a quantized radiation field is
\begin{equation}\label{ham}
{\bf H}_\be = T  + \be V(x).
\end{equation}
\cite{Hi} guarantees the self-adjointness of ${\bf H}_\be$ on the domain
${\mathcal {D}}(p^2 + H_f)$. Our
goal is to show that the operator ${\bf H}_{\be}$ has a  bound
state for  values of $\be$ strictly smaller than $\be_0$. To establish the
existence of a ground state of ${\bf H}_\be$ we apply the
criterion of \cite{GLL}, which says that  ${\bf H}_\be$ has a
ground state if
\begin{equation}\label{crite}
\infspec {\bf H}_{\be} <  \infspec T.
\end{equation}
However, in contrast to the Schr\"odinger operator $h_\be$, for
which the infimum of the spectrum without potential is always $0$,
the $\infspec T$ is a complicated function depending on $\al$ and
$\Lambda$. To prove the inequality (\ref{crite}) one needs precise
estimates on this function. Our first result is the following
asymptotic estimate on $\so = \infspec T$.

\begin{thm}[Localization of the spectrum of a free spinless particle]\label{pt2}
Let
\begin{equation}\label{c1}
{{\mathcal E}_0} = \langle 0| D\cdot D[P_f^2 + H_f]^{-1} D^*\d D^*|0\rangle,
\end{equation}
with $D^* = D^*(0)$ and $|0\rangle $ denotes the vacuum of $\F$.
Then, for small $\al$,
\begin{equation}\label{coeb}
\Big|\Sigma_\al - \ap \Lambda^2+ \al^2 \Em_0\Big| \leq C\al^3
\la^4,
\end{equation}
where $C >0$ is an appropriate constant independent of $\al$ and $\la$.
\end{thm}
\begin{rem}
The number $\Em_0$ can be computed directly through the integral
\begin{equation}
\Em_0 =  \sum_{\mu,\nu =1,2}2 \int \frac{\big[G^\mu (k_1)\cdot
G^\nu(k_2)\big]^2}{|k_1 + k_2|^2 + |k_1| +|k_2|}dk_1 dk_2.
\end{equation}
\end{rem}

The first to leading order is obtained by perturbation theory. One
of our main goals here is to prove that perturbation theory is
correct in the case when $\la$ is fix, which is a non-trivial problem,
since there is no isolated eigenvalue and Kato's perturbation methods cannot be applied.

Recall, that the operator $h_\be$ has a critical value $\be_0$ of
the parameter $\be$ such that, for $\be\leq \be_0$, $h_\be$ does
not have a bound state. Using Theorem \ref{pt2} we construct a
variational trial function proving for small values of $\al$ the
following:
\begin{thm}[Enhanced binding]\label{mt}
For all   sufficiently small $\al$ there exists a number
$\be_1{(\al)} < \be_0$, such that for all $\be > \be_1{(\al)}$ the
operator ${\bf H}_{\be}$ has a ground state.
\end{thm}

Observe that the converse statement is not proven, namely we do not obtain 
a $\be_2(\al) > 0$ such that for $\be < \be_2(\al)$ the ground state does not exist.
\begin{rem} 
Concerning the critical case $\be = \be_0$ the proof of Theorem \ref{mt}
in particular implies that there exists a real number  $\rho > 0$
 such that  ${\bf H}_{\be_0}$ has a ground state for all
$\al\in (0,\rho]$.
\end{rem}
Of course we expect that binding holds on, or even increases, when
$\al$ gets large, but we cannot prove it due to the fact that we can only
control the self-energy  for small $\al$. 

\section{Proof of theorem \ref{pt2}}

Let us start with  a free spinless electron. In this case the
Hamiltonian  is translation invariant, which means that it
commutes with the total momentum $p+ P_f$.
 It is
therefore possible to rewrite the Hilbert space and the
Hamiltonian as a direct integral
\begin{equation}
\Hh=\int_{\R^3}^\oplus d^3P\, \Hh_P
\end{equation}
and
\begin{equation}
T=\int_{\R^3}^\oplus d^3P\, T_P \ ,
\end{equation}
with $T_P$ acting on $\Hh_P$. Each $\Hh_P$ is isomorphic to $\F$.
In this representation  $T_P$ is given by
\begin{equation}
T_P = \big(P-P_f + \sqrt{\alpha}A(0)\big)^2 +
  H_f\ .
\end{equation}
According to \cite{F} the minimum of $\infspec T_P$ is achieved
for $P=0$, which tells us that we only need to consider the
operator
\begin{equation}
T_0 = \big(P_f + \sqrt{\alpha}A\big)^2 +
  H_f.
\end{equation}
Throughout this section we use  $A = A(0)$, $D=D(0)$, and $D^*=
D^*(0)$. We define $\so= \infspec \T$.

It turns out to be convenient to denote a general $\Psi \in \Hh$
as
\begin{equation}
\Psi = \{\psi_0, \psi_1,\dots, \psi_n,\dots\},
\end{equation}
where
\begin{equation}
\psi_n=\psi_n(x,k_1,\dots,k_n;\lambda_1,\dots,\lambda_n).
\end{equation}

In order not to overburden the paper with too many indices we will
suppress the photon variables in $\psi_n$, when it does not lead
to misunderstanding. 

\subsection{Upper bound}
We take the trial state
\begin{equation}
\Psi = \{\ora, 0, -\al [P_f^2 + H_f]^{-1} D^*\d D^*\ora,0,0...\}
\end{equation}
where $\ora \in \C, \langle 0\ora =1$, denotes the vacuum
vector. The photon part of $\Psi$ can be written explicitly as
\begin{equation}
-\sum_{\lambda, \mu =1,2} \al \sqrt{2} \frac 1{(k_1 + k_2)^2 + |k_1|^2 +
|k_2|} G^\lambda(k_1)\cdot G^\mu(k_2) \ora.
\end{equation}
Since
\begin{equation}
 D\d D^* - D^*\d D = \pi^{-1} \la^2,
\end{equation}
obviously
\begin{equation}\label{aq}
A^2 =  \pi^{-1} \la^2 + 2 D^*\d D + D\d D + D^*\d D^*.
\end{equation}
Therefore, since $\|\Psi\|^2 \geq 1$,
\begin{multline}\label{ubm}
\big(\Psi, \T \Psi\big)/(\Psi,\Psi) \leq \ap \la^2 - \al^2 \langle
0| D\d D[P_f^2 + H_f]^{-1} D^*\d D^*|0\rangle \\ + 2\al^3 \|D[P_f^2 +
H_f]^{-1}D^*\d D^*\ora\|^2.
\end{multline}
One can easily see by scaling that the last two terms in the
r.h.s. of (\ref{ubm}) are of order $\la^2$.

\subsection{Lower bound$^1$}

\footnotetext[1]{
A different  proof of the lower bound, based on partitions of unity of the photon configuration space and 
improved estimates for the localization errors for the relativistic energy, can
be found in the preprint version \cite{HVV}.}

We start with some a priori estimates.
\begin{lem}\label{lbl}
\begin{equation}
\T \geq \al \pi^{-1}\la^2 - \const \al^2 \la^3 +\half (P_f^2 + H_f).
\end{equation}
\end{lem}
\begin{proof}
Since $[P_f,A]=0$
\begin{equation}
\T = P_f^2 + 2\as P_f\d A + \al A^2 + H_f.
\end{equation}
By means of Schwarz's inequality
\begin{equation}
2\as P_f\cdot A = 4\as \Re ( P_f\d D) \leq \half P_f^2 + 8\al D^*D
\end{equation}
and
\begin{equation}
\al(D\d D + D^*\d D^*) \leq C^{-1} D^*\d D + \al^2 C D\d D^*
\end{equation}
for any $C > 0$. Using (\ref{aq}) we obtain
\begin{multline}
\T \geq (\pi^{-1} \al \la^2 - C\pi \al^2 \la^2 ) + H_f \Big(\half
- \frac{8\al\la}{\pi} - C^{-1}\la \frac 2\pi - \al^2\frac C\pi
\la\Big)\\ + \half (P_f^2 + H_f),
\end{multline}
which implies the lemma with $C = \bar c \la$, with an appropriate $\bar c>0$,  and $\al \la$ and
$\al$ not too large.
\end{proof}

\begin{rem}
We know
from  the upper bound  that any approximate ground  state $\po$ satisfies
$(\po,\T\po) \leq \al \pi^{-1} \la^2 + \Ow(\al^2)$. Therefore by
Lemma \ref{lbl} we infer the a priori estimate
\begin{equation}\label{apri}
\big(\po,[P_f^2 + H_f]\po\big) \leq \const \al^2 \la^3.
\end{equation}
\end{rem}

Using (\ref{aq}) we derive
\begin{equation}
(\po,T_0 \po) \geq \ap \Lambda^2 \pa \Psi_0 \pa^2 + \sum_{n
=0}^{\infty} \Em[\psi_n,\psi_{n+1},\psi_{n+2}],
\end{equation}
where
\begin{multline}
\Em[\psi_n,\psi_{n+1},\psi_{n+2}] = (\psi_{n+2} , \mA \psi_{n+2})
\\ + 2\Re\left(\big[2\as P_f \d\Da\psi_{n+1}
+ \al \Da\d\Da \psi_n\big], \psi_{n+2}\right),
\end{multline}
with
\begin{equation}
\mA= P_f^2 + H_f.
\end{equation}
Recall, in our notation
\begin{equation}
\Psi_0 = \{\psi_0, \psi_1(k_1),....,\psi_n(k_1,\dots,k_n),...\}.
\end{equation}

We consider the term $\Em[\psi_n,\psi_{n+1},\psi_{n+2}]$. If we
set
\begin{equation}
f= \mA^{1/2} \psi_{n+2}, \,\, g=-\mA^{-1/2}\big[\as 2P_f\d
D^*\psi_{n+1} + \al D^*\d D^*\psi_n\big]
\end{equation}
then by means of $\|f\|^2 - 2\Re(f,g)\geq -\|g\|^2$ we derive
\begin{multline}\label{schw}
\Em[\psi_n, \psi_{n+1}, \psi_{n+2}] \geq- \Big\|2 \as
\mA^{-1/2}P_f\d \Da \psi_{n+1} +\al \mA^{-1/2} \Da\d \Da \psi_n
\Big\|^2.
\end{multline}
Let us start the estimation of the  r.h.s. of (\ref{schw}) with
\begin{equation}\label{ddd}
-\al^2( \psi_n, D\d D[P_f^2 + H_f]^{-1} D^*\d D^*\psi_n).
\end{equation}
It will be shown that it produces the first to leading order in
$\al^2$. Recall,
\begin{multline}\label{defDD}
[D^*\d D^*\psi_n]_{n+2} = \frac 1{\sqrt{(n+2)(n+1)}}
\sum_{\lambda,\mu =1,2} \sum_{j=1}^{n+2} \sum_{\substack{i=1 \\ i \neq j}}^{n+2}
G^\mu(k_j)\cdot G^\lambda(k_i) \times\\\times
\psi_n(k_1,\dots,\not\!\! k_j,\dots,\not \!\! k_i,\dots,k_{n+2}),
\end{multline}
where $\not \!\! k_j$ indicates that the $j-$th variable is
omitted. By permutational symmetry we can distinguish between
three different terms,
\begin{equation}\label{defI}
\big( \psi_n, D\d D [P_f^2 + H_f]^{-1} D^*\d D^* \psi_n \big)=
I_n+II_n+III_n,
\end{equation}
which come out quite naturally when we insert equation (\ref{defDD}) into
(\ref{defI}) and 
have in mind that the l.h.s. of (\ref{defI}) can be written as 
\begin{equation}\label{defDD2}
\big( D^*\d D^* \psi_n,  [P_f^2 + H_f]^{-1} D^*\d D^* \psi_n \big).
\end{equation}

First, the diagonal part $I_n$ appears, when in the right hand side of
(\ref{defDD2}) as well as in the left hand side two photons $G^\mu(k_j)\cdot
G^\lambda(k_i)$ with the same
variables $k_i,k_j$ are produced,
\begin{equation}
I_n= \sum_{\lambda,\mu=1,2} 2 \int \frac{\big[G^\lambda(k_1)\cdot
G^\mu(k_2)\big]^2
|\psi_n(k_3,\dots,k_{n+2})|^2}{\big|\sum_{i=1}^{n+2}k_i\big|^2 +
\sum_{i=1}^{n+2}|k_i|}  dk_1\dots dk_{n+2}.
\end{equation}
If we set $\Q =\big|\sum_{i=3}^{n+2}k_i\big|^2 + \big|k_1 +
k_2\big|^2+ \sum_{i=1}^{n+2}|k_i|$ and $b=
2\big[\sum_{i=3}^{n+2}k_i\big]\cdot \big[k_1+k_2\big]$ and use the
expansion
\begin{equation}\label{exp}
\frac 1{\Q+ b} = \frac 1{\Q} - \frac 1{\Q}b\frac 1{\Q} +  \frac
1{\Q}b\frac 1{\Q+b}b\frac 1{\Q}
\end{equation}
then we see that the second term vanishes when integrating over
$k_1,k_2$. Therefore, with $\Q \geq \big|k_1 + k_2\big|^2+ |k_1| +
|k_2|$ and $\Q+b \geq |k_1| + |k_2|$ we arrive at
\begin{multline}
I_n\leq  \sum_{\lambda,\mu =1,2}2 \Big[ \|\psi_n\|^2\int
\frac{\big[G^\lambda(k_1)\cdot G^\mu(k_2)\big]^2}{|k_1 + k_2|^2 +
|k_1| + |k_2|} dk_1 dk_2 \\ + 4\int
\frac{\big|G^\lambda(k_1)\big|^2\big| G^\mu(k_2)\big|^2 \big[|k_1|
+ |k_2|\big]^2}{\big[|k_1 + k_2|^2 + |k_1| +
|k_2|\big]^2(|k_1|+|k_2|)}\times \\ \times
\big|\sum_{i=3}^{n+2}k_i\big|^2|\psi_n(k_3,\dots,k_{n+2})|^2dk_1\dots
dk_{n+2}\Big] \\ \leq  \langle 0| D\d D[P_f^2 + H_f]^{-1} D^*\d D^*\ora
\|\psi_n\|^2 +\const \la \|P_f\psi_n\|^2.
\end{multline}

For convenience we define the operator $|D|$ by
\begin{equation}
|D| =\sum_{\lambda=1,2} \int |G^\lambda(k)| a_\lambda(k) dk.
\end{equation}
$|D|^*$ denotes the operator adjoint. Obviously, \cite[Lemma A.
4]{GLL} still holds for  $|D|$, namely
\begin{equation}\label{auxeq}
|D|^*|D| \leq \frac 2\pi H_f.
\end{equation}
The second term $II_n$ occurs, when a term $G^\mu(k_j)\cdot
G^\lambda(k_i)$ in the l.h.s. of (\ref{defDD2}) meets a two photon part $G^\mu(k_j)\cdot
G^\lambda(k_l)$ in the r.h.s. of (\ref{defDD2}). Using $P_f^2 + H_f \geq H_f$ we 
 evaluate
\begin{multline}
II_n \leq (n+1)\sum_{\lambda,\mu=1,2}
\int\frac{\big|G^\lambda(k_1)\big|\big|
G^\mu(k_2)\big|\big|G^\lambda(k_1)\big|\big|
G^\mu(k_{n+2})\big|}{\sum_{i=1}^{n+2} |k_i|} \times \\ \times
|\psi_n(k_3,\dots,k_{n+2})||\psi_n(k_2,\dots,k_{n+1})| dk_1\dots
dk_{n+2}\\ \leq \const \int \frac{|G(k_1)|^2}{|k_1|}dk_1
\big(\psi_n,|D|^*|D|\psi_n\big)\leq \const
\la^2\big(\psi_n,H_f\psi_n\big).
\end{multline}

Finally, the third term, where the indices of produced photons in
the right hand side
 differ completely from the indices in the left hand side of (\ref{defDD2}),
can be bounded by
\begin{multline}
III_n \leq (n+1)^2\sum_{\lambda,\mu=1,2}
\int\frac{\big|G^\lambda(k_1)\big|\big|
G^\mu(k_2)\big|\big|G^\lambda(k_{n+1})\big| \big|
G^\mu(k_{n+2})\big|}{\sum_{i=1}^{n+2} |k_i|} \times \\ \times
|\psi_n(k_3,\dots,k_{n+2})||\psi(k_1,\dots,k_{n})| dk_1\dots
dk_{n+2}\\ \leq \const \big(\psi_n,|D|^*H_f^{-1/2}
|D|^*|D|H_f^{-1/2}|D|\psi_n\big) \leq \const \la
\big(\psi_n,H_f\psi_n\big),
\end{multline}
where we used
\begin{equation}
\sum_{i=1}^{n+2} |k_i| \geq \left|\sum_{i=1}^{n+1} |k_i|
\right|^{1/2}\left| \sum_{i=2}^{n+2} |k_i|\right|^{1/2},
\end{equation}
the fact that we can write
\begin{equation}
\big[H_f\big]^{-1/2} \psi_n(k_1,\dots,k_n) = \left[\sum_{i=1}^n
|k_i|\right]^{-1/2} \psi_n(k_1,\dots,k_n),
\end{equation}
and  (\ref{auxeq}).

We summarize
\begin{multline}\label{diego}
-\al^2\big(\psi_n, D\d D \mA^{-1} \Da \d\Da \psi_n\big)\geq -
\al^2\langle 0|D\d D[P_f^2 + H_f]^{-1}\Da\d\Da\ora \|\psi_n\|^2 \\-
\const \la \Big( \|P_f\psi_n\|^2 + \la(\psi_n,H_f\psi_n)\Big).
\end{multline}
The second  diagonal term of (\ref{schw}) reads
\begin{multline}\label{mc}
 -\al (P_f\d\Da \pe, \mA^{-1} P_f\d \Da \pe) =
\sum_{\lambda =1,2} -\al \times \\ \times \Big[ \int \frac{
\left[G^\lambda(k_{n+2}) \cdot \big( \sum_{i=1}^{n+1} k_i\big )
\right]^2 |\pe(k_1,\dots,k_{n+1})|^2}{\left| \sum_{i=1}^{n+2}
k_i\right|^2  + \sum_{i=1}^{n+2} |k_i|}dk_1\dots dk_{n+2}  \\ +
(n+1) \int \frac{ \left[G^\lambda(k_1) \cdot \left(
\sum_{i=1}^{n+2} k_i\right) \right] \left[G^\lambda(k_{n+2}) \cdot
\left(\sum_{i=1}^{n+2} k_i\right)\right]}{\left| \sum_{i=1}^{n+2}
k_i\right|^2 +\sum_{i=2}^{n+2} |k_i|} \times \\ \times
\overline{\pe(k_1,\dots,k_{n+1})}\pe(k_2,\dots,k_{n+2}) dk_1\dots
dk_{n+2}\Big] \\ \geq -\const \al\Big(\la \|P_f\pe\|^2
+(\pe,|D|^*|D|\pe)\Big).
\end{multline}
For the second term in the r.h.s. we used first
\begin{equation}
\frac{\left| \sum_{i=1}^{n+2} k_i\right|^2}{\left|
\sum_{i=1}^{n+2} k_i\right|^2 +\sum_{i=2}^{n+2} |k_i|} \leq 1.
\end{equation}
By (\ref{auxeq}) and Schwarz's inequality for the off-diagonal term
in (\ref{schw}), as well as summing over all $n$ and using the a
priori knowledge (\ref{apri}) we arrive at the desired result.

\section{Proof of theorem \ref{mt}}

To prove the Theorem we will check the binding condition of
\cite{GLL} for $\be = \be_0$. Namely, we will show that
\begin{equation} \label{bindcon}
\infspec {\bf H}_{\be_0} < \so - \delta \al^2 + \Ow(\al^{5/2}).
\end{equation}
The binding for all $\be \in (\be_1,\be_0] $ with some $\be_1 <
\be_0$ follows from (\ref{bindcon}) and the continuity of the
quadratic form in $\be$. In the proof of  Theorem \ref{pt2} we
have seen that the trial state
\begin{equation}
\Psi_n = \{\ora,0, \al [P_f^2 + H_f]^{-1}D(0)^*\d D(0)^*\ora,0,0,..\},
\end{equation}
recovers the self energy  up to the order $\al^2$.
Our next goal is to modify this trial state in such a way that for
the modified state $\Psi^0 \in \Hh$
\begin{equation}
(\Psi^0 , {\bf H}_{\be_0} \Psi^0) \leq (\so - \delta \al^2 +
\Ow(\al^{5/2}))\pa \Psi\pa^2,
\end{equation}
with some $\delta > 0$.

Throughout the previous section we worked with the operator
$A(0)$. Here, our Hamiltonian depends on the electron variable
$x$. In order to adapt our methods developed in the previous
section we introduce the unitary transform
\begin{equation}
U= e^{i P_f\cdot x}
\end{equation}
acting on the Hilbert space $\Hh$. Applied to a $ n$-photon
function $\varphi_n$ we obtain $U\varphi_n= e^{i(\sum_{i=1}^n
k_i)\cdot x} \varphi_n(x,k_1,\dots,k_n)$ and additionally $U
(D^*(x)\psi(x)) = G(k)\psi(x)$.

Since $U p U^* = p - P_f$ we infer for our Hamiltonian ${\bf
H}_{\be_0}$
\begin{equation}
U {\bf H}_{\be_0} U^* = (p - P_f + \as A)^2 +  H_f + \be_0 V(x) \equiv {\bf
\bar H}_{\be_0},
\end{equation}
with $A= A(0)$. Obviously,
\begin{equation}
\infspec {\bf \bar H}_{\be_0}=\infspec {\bf  H}_{\be_0}.
\end{equation}
Therefore, for convenience, we will work in the following with the operator $ {\bf
\bar H}_{\be_0}$.

Next, we define our trial function
\begin{equation} \label{sep}
\Psi^0 = \{f ,  - d \as  {\mA}^{-1} p\d\Da
f, - \al {\mA}^{-1} D^*\d D^*f,
0,0,...\},
\end{equation}
with $\mA = P_f^2 + H_f$, $D = D(0)$, and  $d$ an appropriate
constant which  will be chosen later.

We assume  $f(x)$ $\in C^2_0(\R^3)$ to be a real, spherically symmetric function
and to fulfill the condition
\begin{equation}\label{cond}
\pa p^2 f(x)\pa \leq C_1 \pa p f(x)\pa \leq C_2 \as \pa f(x) \pa,
\end{equation}
with some constants $C_{1,2}$.

For short, denote the 1- and 2- photon terms in $\Psi^0$ as
$\psi_1$ respectively $\psi_2$. Obviously, the terms $(\psi_1, P_f
\cdot p \psi_1)$ and $(\psi_2, P_f \cdot p \psi_2)$ vanish. This
can be seen by integrating over the field variables having in mind
that the reflection $k \to -k$ commutes with $\mA$.

By means of (\ref{cond}) and Schwarz's inequality we obtain
\begin{equation}
\left| 2\as( p\d\Da \psi_1, \psi_2)\right|  +  |(\psi_2,p_x^2
\psi_2)|\leq   \pa \Psi^0\pa^2 \Ow(\al^{5/2}).
\end{equation}
We now use our knowledge from the proof of Theorem 1 to obtain
\begin{multline}
\ap \la^2 \|\Psi^0\|^2 + (\psi_2,[P_f^2 + H_f]\psi_2) + 2\al
\Re (D^*\d D^* f,\psi_2) =\\= \big[\so + \Ow(\al^3)\big]
\|\Psi^0\|^2.
\end{multline}

Taking into account that $V\leq 0$ we arrive at
\begin{multline}\label{seppi}
\big(\Psi^0, {\bf \bar H}_{\be_0} \Psi^0\big) \leq
(f,[p^2 + \be_0V]f) - d \al (f, p\d D
{\mA}^{-1}p\d\Da f)+ \\ + \al d^2\left[ (f,
p\d D{\mA}^{-1} p\d\Da f) + (f, p\d D{\mA}^{-1} p^2
{\mA}^{-1}p\d \Da f)\right] + \\ + [\so + \Ow(\al^{5/2})]
\pa f\pa^2.
\end{multline}
Using the Fourier transform we are able to evaluate explicitly
\begin{equation}
(f,p\d D{\mA}^{-1}p\d \Da f) = \sum_{\lambda =1,2} \int
|\hat f(l)|^2 \frac{[G^\lambda(k) \cdot l]^2}{|k|^2 + |k|}
dk dl =C_3 \pa p f\pa^2
\end{equation}
and additionally get
\begin{equation}
(f,p\d D{\mA}^{-1}p^2{\mA}^{-1}p\d \Da f) =
C_3 \pa p^2 f\pa^2 \leq
C_4 \pa p f\pa^2,
\end{equation}
where we used (\ref{cond}).
This implies
\begin{multline}
\big(\Psi^0, {\bf \bar H}_{\be_0} \Psi^0\big)\leq
\big(1- C_3 \al d + d^2\al(C_3 + C_4)\big)(f,p^2 f)+ (f, \be_0V f)\\ + [\so + \Ow(\al^{5/2})] \pa \Psi^0\pa^2.
\end{multline}

As the next step
we choose $d < \frac{C_3}{2(C_3+C_4)}$ which gives
\begin{equation}\label{2.13}
(\Psi^0 , {\bf\bar H}_{\be_0} \Psi^0) \leq (1 -\nu \al) \pa p f\pa^2 +
\be_0 (f,Vf) + \big(\so  + \Ow(\al^{5/2})\big)\pa \Psi_0\pa^2,
\end{equation}
where
\begin{equation}
\nu = \frac{C_3^2}{4(C_3 +C_4)}.
\end{equation}
Due to our choice of $\be_0$, obviously the operator
\begin{equation}\label{2.14}
-(1-\nu \al)\Delta + \be_0 V(x)
\end{equation}
has at least one negative eigenvalue. However, the r.h.s. of (\ref{2.13}) contains the terms which are
of order $\Ow(\al^{5/2})$
and to prove Theorem \ref{mt} we have to provide more precise estimates on the
negative eigenvalues of (\ref{2.14}).
The required estimate is given by Lemma 2 in the Appendix.

Applying this  Lemma with  $\nu \al = \gamma$ completes the proof of the theorem.

\begin{appendix}

\section{Auxiliary Lemma}

\begin{lem}\label{alem}
Let $\be_0$ be the critical value (the maximal value of the constant $\be$, for which the Schr\"odinger operator
with the potential $\be V$ does not have a discrete spectrum). Then
\begin{equation}
\infspec \{-(1-\gamma)\Delta + \be_0 V(|x|)\} < - \delta \gamma^2,
\end{equation}
for some $\delta >0$ and $\gamma$ small enough.

Moreover,
there exists a function $f_\gamma(x)$, real, spherically symmetric and satisfying condition (\ref{cond}),
with constants $C_{1,2}$ independent of $\gamma$, such that
\begin{equation}
(1-\gamma)\pa \nabla f_\gamma\pa^2 + \be_0 (f_\gamma,  V(|x|)f_\gamma) < - \delta \gamma^2\pa f_\gamma \pa^2.
\end{equation}
\end{lem}

\begin{proof}
Let us start by recalling some properties of the operator
\begin{equation}
h_\be = -\Delta + \be V(x),
\end{equation}
$V(x)\leq 0$, radial, and compactly supported, with critical value $\be =\be_0$. For $\be =\be_0$ the
operator $h_\be$ has a so-called virtual level or zero-resonance. It means that the equation
\begin{equation}\label{app1}
-\Delta \psi + \be_0 V(x) \psi = 0
\end{equation}
has a generalized spherically symmetric solution $\tilde \psi$ with the following properties \cite{VZ}:

\begin{enumerate}
\item[(i)]
Let ${\mathcal B}$ be a closure of the space $C_0^\infty (\R^3)$ in the norm $\pa \psi \pa_B = \pa
\nabla
\psi \pa $. Then $\tilde \psi \in {\mathcal B}$. From this point we assume that $\tilde \psi$ is a
normalized
solution in the sense that $\pa \tilde \psi\pa_B =1$. Notice, that $\tilde \psi \in \Ll^2_{\rm
loc}(\R^3)$,
but $\tilde \psi \notin \Ll^2(\R^3)$.

\item[(ii)]
$-\Delta \tilde \psi \in \Ll^2(\R^3)$ and $V(x)\tilde \psi \in \Ll^2(\R^3)$.

\item[(iii)]
Outside the support of $V(x)$ holds
\begin{equation}\label{a44}
\tilde \psi(x) = C|x|^{-1}.
\end{equation}
\end{enumerate}
The last property follows immediately from the fact that outside the support of
$V(x)$  a radial solution of (\ref{app1}) can be written as $c_1 |x|^{-1} + c_2$,
and $\tilde \psi \in {\mathcal B}$ implies  $c_2 =0$.

Now we proceed directly to the proof Lemma \ref{alem}.
Let
\begin{equation}
u \in C_0^2 (\R^3), \ u(x) \leq 1, \ u(x) =1 \, \, {\rm for} \,\,|x|\leq 1, \,\, u(x) =0
\, \, {\rm for} \,\,|x|\geq 2
\end{equation}
and set
\begin{equation}
f_n(x) = \tilde \psi (x) u(|x|  \gamma n^{-1}) \pa \tilde\psi (x)u(|x|  \gamma n^{-1})\pa^{-1}.
\end{equation}
Obviously $\pa f_n (x)\pa =1$ and for large $n$
\begin{multline}\label{app2}
\pa \nabla f_n(x)\pa \leq \pa \tilde\psi (x)u(|x|  \gamma n^{-1})\pa^{-1}\{\pa \nabla \tilde \psi\pa_{|x|\leq
2\gamma^{-1} n} \\ +
\pa C|x|^{-1}\pa_{\gamma^{-1}n\leq |x| \leq 2\gamma^{-1} n} \max [|\nabla u(|x|\gamma n^{-1})|]\}\\ \leq \pa  \tilde\psi (x)u(|x|
\gamma n^{-1})\pa^{-1}\{\pa \nabla \tilde \psi\pa + c (\gamma^{-1} n)^{1/2}\gamma
n^{-1}\}
\leq 2 \pa\tilde \psi(x)u(|x|  \gamma n^{-1})\pa^{-1}.
\end{multline}

Assume $V(x)$ is supported in a ball of radius $a_0$. Then $\tilde \psi(x) = C|x|^{-1}$ for $ |x| \geq
a_0$ and
\begin{multline}\label{app3}
\pa \tilde \psi (x)u(|x|  \gamma n^{-1})\pa^2 \geq 4\pi C^2\int_{a_0 \leq |x| \leq 2\gamma^{-1}n} d|x| \\ = 4\pi C^2(2\gamma^{-1} n - a_0) \geq
4C^2\pi \gamma^{-1} n
\end{multline}
for $n \geq \frac{a_0}\gamma$.

The inequalities (\ref{app2}) and (\ref{app3}) imply the second relation
in (\ref{cond}) with the constant $C_2$ independent of $n$ and $\gamma$. To check the first inequality in (\ref{cond}) let us estimate
\begin{multline}\label{app4}
\pa \Delta  \left(\tilde \psi (x)u(|x|  \gamma n^{-1})\right)\pa \leq \pa \Delta \tilde \psi\pa + \left( \sum_{i=1}^3
\int \left (\sum_{i=1}^3
\frac{\partial \tilde \psi}{\partial x_i}
\frac{\partial u}{\partial x_i}\right)^2 dx \right)^{\frac 12} \\+ \left(\int \frac{C}{|x|^2} \Delta u(|x|\gamma
n^{-1}) dx\right)^{\frac 12}.
\end{multline}
According to (ii) the first term on the r.h.s. of (\ref{app4}) is bounded. The second term is also
bounded,
since $|\nabla u(|x|\gamma n^{-1})| \leq$ const., (recall that $\pa \nabla \tilde \psi\pa =1$) and the
last term
is also bounded by a constant for $n \geq \frac{2a_0}\gamma$. Finally we arrive at
\begin{equation}
\pa \Delta \left(\tilde \psi (x) u(|x|\gamma n^{-1})\right)\pa \leq C_1\pa \nabla \tilde \psi (x)\pa,
\end{equation}
which implies
\begin{equation}
\pa \Delta f_n (x) \pa \leq C_1\pa \nabla f_n(x)\pa .
\end{equation}

To prove the Lemma it suffices now to show that for large $n$
\begin{equation}
(1-\gamma)\pa \nabla f_n\pa^2 + \be_0(f_n,Vf_n)\leq - \delta \gamma^2\pa f_n\pa^2,
\end{equation}
with some $\delta >0$ independent of $\gamma$. This is equivalent to
\begin{equation}
(1- \gamma)\pa \nabla (\tilde \psi (x)u(|x|  \gamma n^{-1}))\pa^2 + \be_0(\tilde \psi,V\tilde \psi)\leq -
\delta
\gamma^2\pa \tilde \psi (x)u(|x|  \gamma n^{-1})\pa^2.
\end{equation}
Recall that
\begin{equation}\label{sex}
\pa \tilde \psi u(|x|\gamma n^{-1})\pa^2 \leq c_3  \frac 43 \pi a_0^3+ 8\pi C^2 \int_{a_0}^{2\gamma^{-1}n} d|x| \leq 2c_4 \gamma^{-1}n,
\end{equation}
for large $n$, where $c_3 = \max_{|x| \leq a_0} |\tilde \psi (x)|$, and
\begin{equation}
\pa \nabla \tilde \psi\pa^2 + \be_0(\tilde \psi,V\tilde\psi) = 0,
\end{equation}
which implies
\begin{eqnarray}\nonumber
&&(1- \gamma)\pa \nabla (\tilde \psi (x)u(|x|  \gamma n^{-1}))\pa^2 + \be_0(\tilde \psi,V\tilde \psi)\leq \\
\label{app13}
&& \quad \leq -\gamma [\pa (\nabla \tilde \psi)u\pa - \pa \tilde \psi \nabla u \pa]^2 +
[\pa\nabla \tilde \psi\pa^2 - \pa \nabla (\tilde \psi u)\pa^2]\\\nonumber
&& \quad\leq - \gamma \left[ \frac 12 \pa \nabla \tilde \psi\pa - C\gamma^{1/2} n^{-1/2} \right]^2 +
3\pa \nabla \tilde \psi\pa^2_{|x| \geq \gamma^{-1} n} \\ \nonumber && \qquad+ 2\pa \tilde \psi |\nabla u|\pa^2_{\gamma^{-1} n \leq
|x| \leq 2 \gamma^{-1} n}.
\end{eqnarray}
For $n$ large we have
\begin{eqnarray}\nonumber
&i)& \pa \nabla \tilde \psi\pa^2_{|x| \geq \gamma^{-1} n} =4\pi C^2
\int_{\gamma^{-1} n}^\infty |x|^{-2} d|x| = 4\pi C^2
\gamma n^{-1},\\
&ii)& \pa \tilde \psi |\nabla u|\pa^2_{\gamma^{-1} n\leq |x| \leq 2 \gamma^{-1} n} \leq c_5 \gamma^2 n^{-2}
4\pi\int_{\gamma^{-1} n}^{2\gamma^{-1} n}  d|x|
= c_6 \gamma n^{-1},\\ \label{app14}
&iii)& C\gamma^{1/2} n^{-1/2} < \frac 14 = \frac 14 \pa \nabla \tilde \psi\pa,
\end{eqnarray}
which implies together with (\ref{app13})
\begin{multline}\label{irland}
- (1- \gamma)\pa \nabla (\tilde \psi (x)u(|x|  \gamma n^{-1}))\pa^2 + \be_0(\tilde \psi,V\tilde \psi) \\ \leq
- \gamma/4 + 3 C^2\gamma n^{-1} + 2 c_6 \gamma n^{-1} \leq -\gamma/8 \leq -\frac {\gamma^2}{32C^2\pi n} \pa \tilde \psi u(|x|  \gamma n^{-1}))\pa^2.
\end{multline}
To complete the proof of the Lemma it suffices now to choose $n$ so large that (\ref{irland}) holds  true
(notice, that it can be done uniformly in
$\gamma$ for $\gamma \leq 1$) and for this $n$ take $\delta =\frac 1{32} C^{-2} \pi^{-1}n^{-1}$, where $C$ is the
constant in (\ref{app3}),
which depends on the zero-resonance solution $\tilde\psi$ only.
\end{proof}

\end{appendix}
\bigskip

\noindent {\it Acknowledgment:} The work was partially supported
by the European Union through its Training, Research, and Mobility
program FMRX-CT 96-0001. C. Hainzl has been supported by a Marie
Curie Fellowship of the European Community programme \lq\lq Improving Human
Research Potential and the
Socio-economic Knowledge Base\rq\rq\ under contract number
HPMFCT-2000-00660. C. H. and S.-A. V.  thank Robert Seiringer for
many valuable comments.

\end{document}